\documentclass[12pt]{article}
\usepackage{amsmath}
\usepackage{graphicx,psfrag,epsf}
\usepackage{enumerate}
\usepackage{url} 
\usepackage{bm}
\usepackage{verbatim, color, array}

\usepackage{amsthm, amssymb}
\usepackage{float}

\usepackage{hhline}
\usepackage{booktabs}
\usepackage{soul}
\usepackage[super]{natbib}
\usepackage[table]{xcolor} 
\usepackage{hyperref}
\usepackage{multirow}
\usepackage{nicematrix}
\usepackage{threeparttable}
\usepackage{enumitem}
\usepackage{tikz}
\usepackage{appendix}
\usetikzlibrary{shapes, arrows, positioning, calc, shadows, fit}
\usepackage{longtable}
\usepackage[affil-it]{authblk} 
\usepackage{threeparttable}

\newcommand{\blind}{0}

\begin{document}

\def\spacingset#1{\renewcommand{\baselinestretch}%
{#1}\small\normalsize} \spacingset{1}


\if0\blind
{
  \title{\bf A Comparative Evaluation of Bayesian Model-Assisted Two-Stage Designs for Phase I/II Clinical Trials}
  \author{Hao Sun\\
    Global Biometrics \& Data Sciences, Bristol Myers Squibb\\
    and \\
    Jerry Li \\
    Global Biometrics \& Data Sciences, Bristol Myers Squibb}
  \maketitle
} \fi

\if1\blind
{
  \bigskip
  \bigskip
  \bigskip
  \begin{center}
    {\LARGE\bf Title}
\end{center}
  \medskip
} \fi

\bigskip
\begin{abstract}
The primary goal of a two-stage Phase I/II trial is to identify the optimal dose for the following large-scale Phase III trial. Recently, Phase I dose-finding designs have shifted from identifying the maximum tolerated dose (MTD) to the optimal biological dose (OBD). Typically, several doses are selected as recommended Phase II doses (RP2D) for further evaluation. In Phase II dose optimization trials, each RP2D is evaluated independently to determine its "go/no-go" decision. The optimal RP2D is then chosen from the remaining RP2Ds as the recommended Phase III dose (RP3D). The effectiveness of both dose-finding and dose optimization designs at two stages impacts RP3D selection. This paper reviews and compares fifteen Bayesian model-assisted two-stage designs, combining five Phase I dose-finding designs (BOIN, TITE-BOIN, BF-BOIN, BOIN12, and TITE-BOIN12) with three Phase II dose optimization designs (TS, BOP2, and TOP). We conduct extensive simulation studies to evaluate their performance under different dose-response scenarios, with and without the existence of the OBD. Based on our results, we recommend the TITE-BOIN12 + TOP combination as the optimal two-stage design for Phase I/II trials.
\end{abstract}

\noindent%
\vfill

\newpage
\spacingset{1.45} 
\section{Introduction}\label{sec:intro}
The primary objective of a two-stage Phase I/II trial is to evaluate the preliminary toxicity and efficacy outcomes across a range of treatment dosages, from which to identify the optimal dose as the recommended Phase III dose (RP3D). Typically, in a two-stage Phase I/II trial, the Phase I part focuses on dose exploration or dose finding. In a conventional Phase I dose-finding design, the goal is to identify the maximum tolerated dose (MTD), defined as the highest dose associated with an acceptable rate of dose-limiting toxicity (DLT). Following the dose exploration, appropriate doses are selected as the recommended Phase II doses (RP2Ds). The Phase II part is dedicated to dose expansion or dose optimization. In the Phase II trial, each patient is randomly assigned to one RP2D. Both toxicity and efficacy data for each RP2D are analyzed to make a ``go/no-go decision". If at least one RP2D meets the criteria for a ``go" decision at the end of Phase II, the optimal RP2D is then selected as the RP3D for further investigation in a large-scale Phase III trial. 


Conventional Phase I designs focus on identifying the MTD  based on the assumption that drug efficacy increases with dose. Under this assumption, the MTD is the dose offering the highest efficacy among safe dose levels. These dose-finding designs can be classified into rule-based, model-based, and model-assisted methods. The rule-based 3+3 design \cite{storer1989design} is the most widely used approach; however, it is often criticized for its lack of precision \cite{chiuzan20243+}. Model-based designs, such as CRM \cite{CRM1990}, EWOC \cite{Babb1998}, and BLRM \cite{neuenschwander2008critical}, utilize statistical models to characterize the dose-toxicity relationship, demonstrating improvements in MTD-finding accuracy compared to the 3+3 design. Model-assisted designs, such as BOIN \cite{liu2015bayesian}, mTPI \cite{ji2013modified}, and Keyboard \cite{yan2017keyboard}, do not require complex models like model-based designs, yet they also provide reliable MTD identification. To address challenges such as fast accrual rates and increased resource demands in clinical trials, many novel dose-finding designs incorporating time-to-event (TITE) toxicity outcomes have been developed, including TITE-CRM \cite{Cheung2000}, TITE-BOIN \cite{yuan2018time}, and TITE-keyboard \cite{lin2020time}. These TITE designs can reduce trial duration without compromising MTD identification accuracy. Furthermore, \citet{zhao2024backfilling} developed the BF-BOIN design, which incorporates a backfilling strategy into BOIN by treating patients at lower doses where efficacy responses have been observed. This backfilling approach enhances patient enrollment and safety without extending the trial duration.

The aforementioned MTD-finding designs, which are well-suited for chemotherapies where efficacy typically increases with dose, may be less appropriate for emerging therapeutic areas such as immunotherapies and molecularly targeted therapies, where efficacy often does not increase accordingly. If the dose-response relationship is non-monotonic, the MTD may not be the optimal dose, since a lower dose could potentially be more efficacious. Therefore, the FDA has launched Project Optimus \citep{fda_project_optimus} which advocates for dose optimization to identify doses that are both safe and efficacious. This shift in focus has led dose-finding trials to prioritize determining the optimal biological dose (OBD) to achieve the best balance of risk and benefit. Various model-based designs have been developed to model both the dose-toxicity and dose-efficacy curves, such as Eff-Tox \cite{thall2004dose}, L-logistic \cite{zang2014adaptive}, CP-logistic \cite{sato2016adaptive}, and B-dynamic \cite{liu2016robust}. However, model-assisted designs are often preferred for OBD identification, as they do not require model assumptions about the dose-efficacy relationship. Examples include BOIN12 \cite{lin2020boin12}, BOIN-ET \cite{takeda2018boin}, TEPI-2 \cite{li2020tepi}, PRINTE \cite{lin2021probability}, and STEIN \cite{lin2017stein}. Additionally, some OBD-finding designs using TITE outcomes have been developed, such as TITE-BOIN12 \cite{zhou2022tite}, TITE-BOIN-ET \cite{takeda2020tite}, LO-TC \cite{jin2014using}, and Joint TITE-CRM \cite{barnett2024joint}. 

Many design methods have been developed for Phase II clinical trials to support the ``go/no-go decision". One of the most widely adopted designs is Simon's two-stage design \cite{simon1989optimal}, a frequentist method that assumes a single binary toxicity or efficacy outcome and incorporates pre-specified interim analyses. This design enables early trial termination if the treatment is considered toxic or inefficacious at an interim analysis. Extensions of Simon's two-stage design include Fleming's multiple-stage test \cite{fleming1982one}, Chen's optimal three-stage design \cite{chen1997optimal}, and Ensigh's optimal three-stage design \cite{ensign1994optimal}. Bayesian designs for Phase II trials include Thall and Simon's design (TS) \cite{thall1994practical}, the Bayesian predictive probability (PP) approach \cite{lee2008predictive}, BOP2 \cite{zhou2017bop2}, and TOP \cite{zhou2020bayesian}. TS and PP are limited to a single outcome, whereas both BOP2 and TOP can handle complex outcomes, with TOP extending BOP2 by incorporating TITE outcomes. \citet{yang2024design} developed the MERIT design which focuses on a multiple-dose randomized trial with sample size determination. 

In recent decades, numerous studies have focused on evaluating the performance of dose-finding designs for Phase I or dose optimization designs for Phase II. However, to the best of our knowledge, no studies have analyzed the combination of a dose-finding design with a dose optimization design and provided recommendations for a two-stage Phase I/II clinical trial. Traditional evaluations of dose-finding designs primarily focused on the probability of selecting the MTD or the OBD. However, after the Phase I part of a two-stage trial, it is common practice to select at least two RP2Ds for the following Phase II part. Therefore, accurately selecting the OBD as an RP2D becomes crucial for the Phase I part of a two-stage Phase I/II design. Additionally, other performance metrics should also be considered in a two-stage Phase I/II trial, such as the probability of selecting at least one toxic dose as an RP2D, the probability of identifying the OBD as the final RP3D, and the probability of early trial termination at each stage if the OBD does not exist. This paper evaluates the performance of various two-stage designs at both the Phase I part and overall, using ten different performance metrics for a comprehensive assessment. Given the limited sample sizes of a two-stage Phase I/II trial, We focus on Bayesian designs for both stages. We consider five dose-finding designs for the Phase I part: BOIN, TITE-BOIN, BF-BOIN, BOIN12, and TITE-BOIN12. BOIN is a widely used MTD-finding design and has been applied in numerical clinical trials. It has been recognized as a ``fit-for-purpose" design by the FDA \cite{us2022drug}. The other four designs are all extended designs of BOIN. Among them, TITE-BOIN and BF-BOIN are MTD-finding designs, while BOIN12 and TITE-BOIN12 are OBD-finding designs. Note that BOIN12 is the only OBD-finding design that has been used in real-world clinical trials, such as in NCT05032599 — a first-in-human (FIH), single-center, open-label, non-randomized, single-arm Phase I trial evaluating the safety and tolerability of CD5 CAR T-cells in patients with relapsed or refractory T-cell acute lymphoblastic leukemia. For the Phase II part, we consider three designs: TS, BOP2, and TOP. Both TS and BOP2 are widely used in Phase II trials. For example, TS has been used in NCT00548756, NCT03734029, and NCT03819985, while BOP2 has been used in NCT03468218, NCT05361551, and NCT03253679. All the considered designs are straightforward to implement. Software for all designs, except TS, is available at \href{https://trialdesign.org/}{www.trialdesign.org}, which provides an interactive and user-friendly interface. Additionally, the public R package ``ph2bye" is available to implement the TS design.


This article is organized as follows: Section~\ref{sec2_method} introduces notations and the design structure. Key concepts of the dose-finding and dose optimization designs are described in Section~\ref{sec3_methodology}. Section~\ref{sec4_simulation} provides extensive simulation studies and practical recommendations for a two-stage Phase I/II trial. Further discussion is provided in Section~\ref{sec6_discussion}.

\section{Methods}\label{sec2_method}
\subsection{Notations}\label{sec_notation}
We assume that both parts of a two-stage Phase I/II trial collect the same binary toxicity outcome $Y_T$ and binary efficacy outcome $Y_E$, where $Y_T = 1$ indicates a DLT and $Y_E = 1$ represents an efficacy response, such as an objective response (OR). The combination of the binary toxicity and efficacy outcomes consists of four possible results: $O_1$= ($Y_T = 0, Y_E = 1$), $O_2$ = ($Y_T = 0, Y_E = 0$), $O_3$ = ($Y_T = 1, Y_E = 1$), and $O_4$ = ($Y_T = 1, Y_E = 0$) \cite{lin2020boin12}. Suppose there are $D_1$ dose levels in the Phase I part, represented by the index set $I = \{1, 2, \ldots, D_1\}$. At the end of Phase I, up to $D_2$ RP2Ds are selected from the set $I$ to proceed to the Phase II part for dose optimization. We define $\pi_{d,i}$ as the true probability of the $i$-th outcome for dose level $d$, so that $p_d = \pi_{d, 3} + \pi_{d, 4}$ and $q_d = \pi_{d, 1} + \pi_{d, 3}$, where $p_d$ and $q_d$ are the probability of DLT and the probability of efficacy response of dose level $d$, respectively. Let $p_T$ be the maximum acceptable toxicity probability and $q_E$ be the minimum acceptable efficacy probability. A dose level $d$ is considered admissible if it satisfies the criteria $p_d \leq p_T$ and $q_d \geq q_E$. The toxicity probabilities are assumed to increase monotonically, i.e., $p_1 < \ldots < p_D$, while the efficacy probabilities follow an unknown dose-response relationship. 

In Phase I, patients are grouped into cohorts, with each cohort consisting patients who enroll in the trial within a specific time frame. Patients within the same cohort are assigned to the same dose level. We assume that the dose escalation starts at the lowest dose level. Let $C$ represent the maximum number of cohorts, and $n_c$ be the cohort size. Therefore, the maximum sample size for Phase I is $N_1 = Cn_c$. In Phase II, patients are randomly assigned to the RP2Ds with equal probability. Denote $M$ as the maximum number of patients allocated to each RP2D, resulting the maximum sample size of the Phase II part equal to $N_2 = D_2M$. At the end of the Phase II part, one optimal dose will be selected from the RP2Ds as the RP3D for the subsequent Phase III trial. 

Let $n^k_{d,i}$ be the number of patients with outcome $O_i$ at dose level $d$ during Phase $k$, where $i = 1, 2, 3, 4$ and $k = 1, 2$. Therefore, we have $n^k_{d,T} = n^k_{d,3} + n^k_{d,4}$ as the number of patients experiencing a DLT, $n^k_{d, E} = n^k_{d, 1} + n^k_{d, 3}$ as the number of patients experiencing an efficacy response, and $n^k_d = \sum_i^4 n^k_{d, i}$ as the total number of patients at dose level $d$ during Phase $k$. We denote $\hat{p}^k_d$ and $\hat{q}^k_d$ as the observed toxicity probability and the observed efficacy probability of dose level $d$ during Phase $k$, where $\hat{p}^k_d= n^k_{d,T}/n^k_d$, and $\hat{q}^k_d = n^k_{d,E}/n^k_d$, respectively. 

\subsection{Design Structure}
The MTD, denoted $d_{MTD}$, is the dose level with the highest acceptable toxicity probability, i.e., 
$$
d_{MTD} = \arg\max_{d} p_dI(p_d\leq p_T). 
$$
If no dose level has an acceptable toxicity probability, the MTD does not exist. On the other side, the OBD is defined as the optimal dose among the admissible dose levels. Following \citet{lin2020boin12}, we define the OBD using the utility score approach. Let $u_i$ be the utility score associated with $O_i$, where the utility score of the best result $O_1$ is $u_1 = 100$ and the utility score of the worst result $O_4$ is $u_4 = 0$. The utility scores for the other two results, $u_2$ and $u_3$, fall within the interval $[0, 100]$. The expected utility $EU_d$ of dose level $d$ is defined as $EU_d = \sum_{i = 1}^4 \pi_{d, i}u_i$. A higher $EU_d$ indicates greater desirability of a dose, reflecting an improved balance between risk and benefit. Therefore, we define the OBD, denoted $d_{OBD}$, as the dose level with the highest $EU_d$ among the admissible doses, i.e. 
$$
d_{OBD} = \arg\max_{d} EU_d I(p_d\leq p_T, q_d \geq q_E). 
$$ 
Specifically, when $u_2 + u_3 = 100$, we have $EU_d = u_2(1-p_d) + u_3q_d$. Moreover, if $u_2 = 0$ and $u_3 = 100$, then $EU_d = 100q_d$, which maximizes the efficacy probability. In this paper, we set $(u_2, u_3) = (40, 60)$ as recommended by \citet{lin2020boin12}. If no dose level is admissible, the OBD does not exist. 

The OBD and the MTD can differ because the MTD is determined based solely on toxicity probability, assuming that the dose-response relationship increases with dose. Under this assumption, the MTD has the highest efficacy probability among all safe doses. However, the MTD may not always be optimal when considering the overall risk-benefit balance. In new therapeutic areas, the dose-response relationship can take different forms, such as plateau or unimodal patterns, causing the OBD to differ from the MTD. Therefore, even if the MTD is correctly identified and selected as an RP2D through a traditional MTD-finding design in Phase I, there is no guarantee that the OBD will also be included as an RP2D. Note that the OBD may not exist even if the MTD exists, where the MTD and all lower dose levels are inefficacious. In such cases, the Phase II trial should not proceed.

The primary objective of a two-stage Phase I/II trial is to identify the OBD as the RP3D when the OBD exists. However, if the OBD is absent, the trial should be terminated as early as possible. To achieve this, an effective two-stage Phase I/II design should have the following properties: 
\begin{itemize}
    \item[1. ] \textbf{OBD Coverage}: If the OBD exists, the dose-finding design in Phase I should have a high probability of selecting the OBD as an RP2D, and the dose optimization design in Phase II should have a high probability of identifying the OBD as the RP3D when the OBD is within the RP2D list. 
    \item[2. ] \textbf{Prompt Termination}: If the OBD does not exist, the two-stage design should have a high probability of terminating the trial without selecting any dose as the RP3D. Ideally, the trial should terminate before progressing to the Phase II part whenever possible.
    \item[3. ] \textbf{Avoiding Inadmissible RP2D}: When the dose list includes inadmissible doses, a robust two-stage design should control the probability of selecting an inadmissible dose as an RP2D, particularly avoiding toxic doses. Both the probability of selecting a toxic dose as an RP2D and the number of patients allocated to toxic doses should be carefully monitored.
\end{itemize}
We will evaluate these properties among the proposed two-stage designs using ten performance metrics through simulation studies in Section~\ref{sec4_simulation}. 

\section{Methodology}\label{sec3_methodology}
As introduced in Section~\ref{sec:intro}, among the five considered Bayesian model-assisted dose-finding designs for Phase I, BOIN \cite{liu2015bayesian}, TITE-BOIN \cite{yuan2018time}, and BF-BOIN \cite{zhao2024backfilling} are MTD-finding designs, whereas BOIN12\cite{lin2020boin12} and TITE-BOIN12 \cite{zhou2022tite} are OBD-finding designs. TITE-BOIN and TITE-BOIN12 incorporate TITE outcomes, which can help reduce the trial duration. Additionally, BF-BOIN includes a backfilling strategy, allowing patients to be assigned concurrently to previously cleared safe doses during dose escalation. For the Phase II part, we consider three Bayesian model-assisted dose optimization designs: TS \cite{thall1994practical}, BOP2 \cite{zhou2017bop2}, and TOP \cite{zhou2020bayesian}. TS employs a fixed probability cutoff throughout the trial, while BOP2 and TOP utilize adaptive stopping cutoffs that depend on the interim sample size. In addition, TOP is an extension of BOP2 by incorporating TITE outcomes. A detailed description of these designs is provided in Sections \ref{stage_1_designs} and \ref{stage_2_designs}.

\subsection{Dose-Finding Designs in Phase I}\label{stage_1_designs}
We first briefly review BOIN, as other considered dose-finding designs are all BOIN-based designs. Let $\phi$ represent the target toxicity probability. Define two cutoff points, $\phi_1$ and $\phi_2$, where $0 < \phi_1 < \phi < \phi_2 < 1$. We consider three hypotheses for the toxicity probability: 
$$
H_{1}^T: p_d = \phi_1 \quad \text{versus} \quad H_2^T: p_d = \phi \quad \text{versus} \quad H_3^T: p_d = \phi_2
$$
To minimize the probability of incorrect classification of the toxicity outcome, we have 
$$
\lambda_e=\log \left(\frac{1-\phi_1}{1-\phi}\right)\big\slash\log \left[\frac{\phi\left(1-\phi_1\right)}{\phi_1(1-\phi)}\right], \quad \lambda_d = \log \left(\frac{1-\phi}{1-\phi_2}\right) \slash \log \left[\frac{\phi_2\left(1-\phi_2\right)}{\phi\left(1-\phi_2\right)}\right], 
$$
where $\lambda_d$ and $\lambda_e$ are the optimal escalation and de-escalation boundaries for BOIN \cite{liu2015bayesian}. By default, we set $\phi = p_T$, $\phi_1 = 0.6p_T$, and $\phi_2 = 1.4p_T$. For example, if $p_T = 0.3$, then $\lambda_e = 0.236$ and $\lambda_d = 0.359$. 

Let $c = 1$ represent the initial cohort index and $d = 1$ be the pre-specified initial dose level. The dose-finding algorithm of BOIN proceeds as follows: 
\begin{enumerate}
    \item Treat the cohort $c$ at the dose level $d$. 
    \item Calculate the observed toxicity probability $\hat{p}_d^1 = n_{d,T}^1/n_d^1$ at the current dose level $d$. 
    \begin{itemize}
        \item[(a)] if $\hat{p}_d^1 \geq \lambda_d$, de-escalate to the next lower dose level $d-1$; 
        \item[(b)] if $\lambda_e < \hat{p}_d^1 < \lambda_d$, stay at the current dose level $d$; 
        \item[(c)] if $\hat{p}_d^1 \leq \lambda_e$, escalate the next higher dose level $d+1$. 
    \end{itemize}
    \item Set $c = c+1$ and update $d$ based on the dosing decision from Step 2. 
    \item Repeat Steps 1 -- 3 until the maximum sample size is reached. 
\end{enumerate} 
Before making the dosing decision for the next cohort, we ensure the safety of the current dose level by applying a safety criterion: if $\Pr(p_d^1 > \lambda_d \mid n_{d,T}^1, n_d^1) > \eta$, where $\eta$ is a pre-specified probability cutoff, we eliminate the current dose level $d$ and all higher doses from the dose list. If no safe dose is available for the next cohort, the trial will be terminated.

Other BOIN-based designs have similar dose-finding algorithms as BOIN. Unlike BOIN, which requires a complete toxicity assessment for all patients in the same cohort, TITE-BOIN can estimate the observed DLT probability, $\hat{p}_{d}^1$, by imputing pending individual toxicity outcomes \cite{yuan2018time}. In this approach, the number of patients with observed DLT outcome, $n_{d,T}^1$, is known, while the number of patients who complete their toxicity assessment without experiencing a DLT is imputed by the effective number of patients, denoted as $\tilde{m}_{d,T}^1$. Therefore, $\hat{p}_d^1$ is estimated by $\tilde{p}_d^1 = n_{d,T}^1/(n_{d,T}^1 + \tilde{m}_{d,T}^1)$. The primary advantage of using a TITE outcome is to reduce the trial duration. When patient accrual is rapid, new patients can be enrolled without waiting for the full toxicity assessment of all enrolled patients. To avoid making risky decisions when a large proportion of patients have pending toxicity outcomes, we implement an accrual suspension criterion: if more than 50\% of patients at the current dose have pending toxicity outcomes, suspend accrual to allow more data to become available.

In contrast, BF-BOIN divides the entire dose-finding process into two components: the dose escalation component and the backfilling component \cite{zhao2024backfilling}. When the toxicity assessment for the current cohort is still in progress, the backfilling component can enroll new patients at doses that have been designated as safe and have shown at least one efficacy response in the dose escalation component. The dosing decision for the next cohort is made after the toxicity assessment for the current cohort is completed and utilizes toxicity data from both the dose escalation and backfilling components. Compared to TITE-BOIN, which shortens the trial duration while maintaining the maximum sample size the same as BOIN, BF-BOIN allows for an increased number of enrolled patients within a trial duration similar to BOIN.

BOIN12 and TITE-BOIN12 consider both toxicity and efficacy outcomes to identify the OBD, balancing the risk-benefit trade-off. Both designs adopt the expected utility score to generate a rank-based desirability score (RDS) table for the dosing decision \cite{lin2020boin12}. Let $N^* = 6$ be a pre-specified sample size cutoff. The dose-finding algorithm for BOIN12 is outlined as follows \cite{lin2020boin12}:
\begin{itemize}
    \item[1. ] Treat the cohort $c$ at the dose level $d$. 
    \item[2. ] Calculate the observed toxicity probability $\hat{p}_d^1 = n_{d,T}^1/n_d^1$ at the current dose level $d$. 
        \begin{itemize}
            \item[(a) ] if $\hat{p}_{d}^1 \geq \lambda_d$, set $d'$ as the next lower dose level $d - 1$; 
            \item[(b) ] if $\lambda_e < \hat{p}_d^1 < \lambda_d$: (i) when $n_d^1 \geq N^*$, select one dose $d'$ from the admissible set $\{d-1, d\}$ with larger RDS; (ii) when $n_d < N^*$, select one dose $d'$ from the admissible set $\{d-1, d, d+1\}$ with the largest RDS; 
            \item[(c) ] if $\hat{p}_d^1 \leq \lambda_e$, select one dose $d'$ from the admissible set $\{d-1, d, d+1\}$ with the largest RDS.
        \end{itemize}
    \item[3. ] Set $c = c+1$ and update $d = d'$ based on the dosing decision from Step 2. 
    \item[4. ] Repeat Steps 1 -- 3 until the maximum sample size is reached. 
\end{itemize}
The BOIN12 dose-finding algorithm relies on toxicity and efficacy information not only for the current dose but also for other dose levels within the admissible set. This approach accounts for the uncertainty in the dose-response relationship. By comparing the RDS across all doses in the admissible set, BOIN12 identifies the most suitable dose with the highest RDS. In BOIN12's dose exploration, both the safety and futility criteria for the current dose are evaluated. The safety criterion is identical to that of the BOIN design. For futility, if $\Pr(q_d^1 < q_E \mid n_{d,E}^1, n_d^1) > \zeta$, where $\zeta$ is a pre-specified probability cutoff, we eliminate the current dose. As an extension of BOIN12, TITE-BOIN12 has a similar dose-finding algorithm by estimating both observed toxicity and efficacy probabilities. TITE-BOIN12 incorporates safety, futility, and accrual suspension criteria during dose exploration. 


\subsection{Dose Optimization Designs in Phase II}\label{stage_2_designs}
For a Phase II dose optimization trial, the three considered designs are all Bayesian sequential monitoring designs. Suppose there are $R$ interim analyses conducted in Phase II. We define $m_r$ as the sample size enrolled of one RP2D before the $r$-th interim analysis, so that $m_1 < \ldots < m_R < m_{R+1} = M$, where $R+1$ denote the final analysis. Consider the following two point hypotheses for the binary toxicity and efficacy outcomes, respectively: 
$$\begin{aligned}
\text{Toxicity}: H_0^T: p_d = p_{Null} \quad \text{vs} \quad H_1^T: p_d = p_{Alt} \\
\text{Efficacy}: H_0^E: q_d = q_{Null} \quad \text{vs} \quad H_1^E: q_d = q_{Alt} \\
\end{aligned}$$
where $p_{NULL}$ is an unacceptable toxicity probability, $p_{Alt}$ is an ideal toxicity probability, $q_{NULL}$ is a futile efficacy probability, and $q_{Alt}$ is an ideal efficacy probability, with $p_{Alt} < p_{Null}$ and $q_{Null} < q_{Alt}$. At each interim analysis or at the end of the trial, a ``go/no-go" decision is determined. A ``go" decision is made if both the null hypotheses of toxicity and efficacy are rejected; otherwise, a ``no-go'' decision is made. At the end of the Phase II trial, the RP3D will be selected as the optimal dose with the highest utility score among the remaining RP2Ds that receive a ``go'' decision at the final analysis. If no RP2D receives a ``go" decision at the final analysis or does not proceed to the final analysis, the trial is considered terminated.  

Let $C^T(m)$ and $C^E(m)$ denote the probability cutoffs for toxicity and efficacy outcomes, respectively, based on $m$ enrolled patients. Each probability cutoff is a function of the sample size $m$. Within the Bayesian framework, at the $r$-th interim look, an RP2D $d$ is eliminated with a ``no-go" decision if 
$$
\Pr(p_d > p_{Null} \mid \text{Data of } m_r \text{ patients}  ) > C^T(m_r) \quad \text{or} \quad \Pr(q_d \leq q_{Null} \mid \text{Data of } m_r \text{ patients}) > C^E(m_r). 
$$
To determine the cutoff values $\{C^T(m_r), C^E(m_r)\}_{r = 1, \ldots, R+1}$, one common strategy is to maximize the design's statistical power while maintaining the type I error rate at the pre-specified level. In the TS design, $C^T(m)$ and $C^E(m)$ are set as fixed values across all analyses, i.e., $C^T(m_r) = C^T_{TS}$ and $C^E(m_r) = C^E_{TS}$, $r = 1, \ldots, R+1$. Furthermore, TS only considers a single binary outcome so that $C^T_{TS}$ and $C^E_{TS}$ are chosen independently.

On the other hand, BOP2 and TOP can be utilized for complex outcomes, such as the combination of binary toxicity and efficacy outcomes. In this case, the combined outcome $O$ at dose level $d$ is assumed to follow a multinomial distribution: 
$$
O \sim \text{Multinom}(\pi_{d,1}, \pi_{d,2}, \pi_{d, 3}, \pi_{d, 4}).
$$
The probability vector $\bm{\pi}_d = (\pi_{d,1}, \pi_{d,2}, \pi_{d,3}, \pi_{d,4})^T$ is assumed to have a Dirichlet prior, $\text{Dir}(a_1, a_2, a_3, a_4)$, where $a_i$'s are hyperparameters satisfying $\sum_{i=1}^4 a_i = 1$. Therefore, the posterior distribution of $\bm{\pi}_d$ given data at the $r$-th interim analysis is
$$
\bm{\pi}_d \mid \text{Data} \sim \text{Dir}(a_1 + n_{d,1,r}^2, \ldots, a_4 + n_{d,4, r}^2),
$$
where $n_{d,i,r}^2$ represents the number of patients with outcome $O_i$ at dose level $d$ prior to the $r$-th interim analysis, with $\sum_{i=1}^4 n_{d,i,r}^2 = m_r$. By utilizing a combined outcome, BOP2 and TOP are well-suited for handling correlated toxicity and efficacy outcomes. They employ a two-parameter power function: $C(m) = 1 - \lambda(m/M)^\gamma$, where both $\lambda$ and $\gamma$ are tuning parameters that can be chosen differently for the toxicity and efficacy outcomes. In both designs, $\{C^T(m_r), C^E(m_r)\}_{r=1,\ldots, R+1}$ are selected jointly to maximize the overall statistical power while controlling the overall Type I error rate. Because TOP incorporates TITE toxicity and efficacy outcomes, the tuning parameters for BOP2 and TOP can be different.

\section{Simulation}\label{sec4_simulation}
\subsection{Simulation Settings}
We presented nine simulation scenarios to compare the fifteen combinations of two-stage designs for a Phase I/II trial. We examined $D_1 = 6$ dose levels in Phase I, from which up to $D_2 = 2$ doses could be selected as RP2Ds for the Phase II part. In Phase I, we set the cohort size at $n_c = 3$ with a maximum of $C = 15$ cohorts, resulting in a maximum sample size of $N_1 = 45$ for dose escalation. In Phase II, up to $M = 40$ patients could be assigned to each RP2D. We considered $R = 3$ interim analyses with $m_r = 10r$, $r = 1, 2, 3$. Toxicity outcomes were assessed at all interim analyses, whereas efficacy outcomes were evaluated only during the second interim analysis. The patient accrual rate was set at 1 patient per 10 days in Phase I and 1 patient per 5 days in Phase II. In both Phase I and Phase II, the toxicity assessment window was 30 days, while the efficacy assessment window was 90 days. We assumed the maximum acceptable toxicity probability $p_T = 0.35$ and the minimum acceptable efficacy probability $q_E = 0.25$. For the safety and futility criteria across all designs, we applied $\eta = 0.95$ and $\xi = 0.9$. 

Among the nine scenarios, we considered various dose-response relationships: an increasing dose-response in scenarios 1, 2, and 3, a unimodel dose-response in scenarios 4, 5, and 6, and a plateau dose-response in scenarios 7, 8, and 9. In scenarios 1, 4, and 7, the OBD and the MTD were identical, while the OBD was lower than the MTD in scenarios 2, 5, and 8. In scenarios 3, 6, and 9, we assumed that the OBD did not exist. Table~\ref{tab:true_simuvalue} presents the true toxicity and efficacy probability values, as well as the utility score for each dose level across these scenarios.

We conducted 1000 independent replications for each scenario. In each replication, if the Phase I design focused on MTD-finding, the selected MTD and the next lower dose were selected as the RP2Ds. For OBD-finding designs, the admissible dose with the highest utility score was chosen as the RP2D, along with either the next higher or lower adjacent dose, depending on their observed utility scores. If both adjacent doses to the selected OBD were eliminated during the trial, the selected optimal dose became the sole RP2D. At the end of the Phase II part, if both RP2Ds met with the ``go" decision, their utility scores were calculated using their Phase II data, and the dose with the higher utility score would be selected as the RP3D. In cases where the utility scores of both doses were identical, the lower dose was chosen as the RP3D.

\subsection{Simulation Results}
To comprehensively evaluate the performance of the two-stage designs, we analyzed these designs both in Phase I and overall. For Phase I, we assessed the following five performance metrics for each dose-finding design: 
\begin{itemize}
    \item $p_{rp2d}$: the probability that the OBD is selected as an RP2D; 
    \item $p_{rp2d, tox}$: the probability that at least one toxic dose level is selected as an RP2D; 
    \item $p_{et, s1}$: the probability of early termination before progressing to Phase II; 
    \item $n_{tox, s1}$: the average number of patients assigned to a toxic dose in Phase I;
    \item $Dur_{s1}$: the average trial duration of Phase I.
\end{itemize}
To evaluate the overall trial performance, we measure five additional performance metrics: 
\begin{itemize}
    \item $p_{rp3d}$: the probability of selecting the OBD as the final RP3D; 
    \item $p_{et}$: the probability of early termination before completing Phase II;
    \item $n_{total}$: the average total number of enrolled patients throughout the trial;
    \item $n_{tox}$: the average number of patients assigned to a toxic dose during the entire trial; 
    \item $Dur$: the average duration of the entire trial. 
\end{itemize}

Table~\ref{tab:stage_1_results} presents some notable findings about Phase I dose-finding designs. In summary, in scenarios 1, 4, and 7, where the OBD and the MTD were the same dose level, MTD-finding designs had a slightly higher $p_{rp2d}$ than OBD-finding designs. However, when the OBD differed from the MTD (scenarios 2, 5, and 8), OBD-finding designs showed a substantially higher $p_{rp2d}$ than MTD-finding designs. For instance, in scenario 5, where the OBD was dose level 3 and the MTD was dose level 5, BOIN12 and TITE-BOIN12 achieved over an 80\%  probability of selecting the OBD as an RP2D, while MTD-finding designs had around a 40\% probability. When the OBD did not exist (scenarios 3, 6, and 9), OBD-finding designs demonstrated a significantly higher probability of terminating the trial in Phase I, resulting in substantial budget savings. Overall, OBD-finding designs had a lower probability of selecting a toxic dose as an RP2D and assigned fewer patients to toxic doses compared to MTD-finding designs. Among the three MTD-finding designs, BF-BOIN had the lowest $p_{rp2d, tox}$. However, due to the additional patients allocated for backfilling, BF-BOIN enrolled the highest number of patients in Phase I.  Since MTD-finding designs considered only toxicity outcomes, they had a shorter average duration compared to OBD-finding designs. TITE-BOIN had the shortest average duration in Phase I, with TITE-BOIN12 reducing the average duration by approximately 20 months compared to BOIN12. 

The simulation results for the entire study are presented in Table ~\ref{tab:overall_results}. Certain performance metrics specific to the Phase II part alone are not included because the performance of a Phase II trial depends on the RP2Ds selected during Phase I. By keeping the same Phase I design, we can compare the performance of the Phase II designs. As shown in Table ~\ref{tab:overall_results}, the TS design had a much higher probability of terminating the trial in Phase II without selecting any dose as the RP3D across all scenarios. Therefore, in scenarios 1, 2, 4, 5, 7, and 8 where the OBD existed, TS showed a significantly lower $p_{rp3d}$ compared to BOP2 and TOP. Moreover, TS had higher $p_{et}$ values, and lower $n_{total}$, $n_{tox}$, and $Dur$ than BOP2 and TOP in all scenarios. BOP2 and TOP exhibited similar performance across most performance metrics, with the exception of trial duration. By utilizing TITE outcomes, TOP achieved a shorter trial duration than BOP2 in all scenarios, without compromising OBD selection performance. 

Based on the simulation results, our recommended Phase I dose-finding design is TITE-BOIN12, which showed a high probability of selecting the OBD as an RP2D when it existed. TITE-BOIN12 effectively terminated the trial in Phase I when no OBD existed. It also demonstrated a low probability of including a toxic dose in the RP2D list and assigned few patients to toxic doses. In addition, TITE-BOIN12 can shorten the trial duration compared to BOIN12. However, if the primary objective of the Phase I study is to identify the MTD, we recommend BF-BOIN because it outperformed BOIN and TITE-BOIN with a higher $p_{rp2d}$ in most scenarios and consistently lower $p_{rp2d, tox}$, benefiting from its backfilling strategy. For Phase II, our recommendation is the TOP design. As an extension of BOP2, TOP provided flexible probability cutoffs at each interim analysis, achieving a higher probability of correctly identifying the OBD as the RP3D compared to the TS design, while having a shorter trial duration than BOP. Therefore, our overall recommendation for a two-stage Phase I/II trial is to use TITE-BOIN12 in Phase I and TOP in Phase II. However, if the Phase I part aims to identify the MTD, we suggest combining BF-BOIN for Phase I with TOP for Phase II.


\section{Discussion}\label{sec6_discussion}

In this article, we compared fifteen Bayesian model-assisted two-stage designs for Phase I/II trials, by combining five Phase I dose-finding designs (BOIN, TITE-BOIN, BF-BOIN, BOIN12, and TITE-BOIN12), with three Phase II dose optimization designs (TS, BOP2, and TOP). Based on our simulation results, BOIN12 and TITE-BOIN12 showed a higher probability of selecting the OBD as an RP2D and a higher probability of terminating the trial in Phase I when the OBD did not exist, compared to the three MTD-finding designs. However, when the OBD and the MTD were the same dose, the MTD-finding designs performed slightly better in OBD selection than the OBD-finding designs. The OBD-finding designs outperformed the MTD-finding designs in overdose control across all scenarios. Specifically, BOIN12 and TITE-BOIN12 had a lower probability of selecting a toxic dose as an RP2D and allocated fewer patients to toxic doses than the other three MTD-finding designs. Among the three Phase II designs, BOP2 and TOP outperformed TS in correctly identifying the OBD as the RP3D and were also less likely to terminate the trial prematurely when the OBD existed. Therefore, we recommend TITE-BOIN12 + TOP as the optimal two-stage design for Phase I/II trials among the fifteen combinations.

This article has several limitations. First, we considered only a limited number of designs for both stages. For example, some other BOIN-based designs, such as BOIN-ET \cite{takeda2018boin} and STEIN \cite{lin2017stein}, were not included in our evaluation. \citet{sun2024statistical} compared seven model-assisted OBD-finding designs (BOIN12, BOIN-ET, TEPI-2, PRINTE, STEIN, UBI, and uTPI), which showed that STEIN outperformed BOIN12 in OBD selection and overdose control across multiple dose-response relationships. However, both STEIN and BOIN-ET were less competitive than BOIN12 when the OBD did not exist. Besides, STEIN has not yet been used in a real trial and lacks public software, whereas BOIN-ET can be implemented using the R package ``boinet". Second, we assumed a fixed sample size for the Phase II part. If achieving a pre-specified statistical power with type I error control is necessary, the three methods we considered may not be suitable, and designs like MERIT \citet{yang2024design} may be more appropriate. Third, our study only considered binary outcomes for toxicity and efficacy. However, BOIN12, TITE-BOIN12, BOP2, and TOP can accommodate ordinal outcomes, although we did not evaluate their performance with ordinal outcomes in our simulation. Several dose-finding designs have been specifically developed for ordinal outcomes, including gBOIN \cite{mu2019gboin}, TITE-gBOIN \cite{takeda2022tite}, gBOIN-ET \cite{takeda2022gboin}, and TITE-gBOIN-ET \cite{takeda2023tite}. Fourth, pharmacokinetics (PK) outcomes are commonly collected in early-phase trials to assess drug exposure and are often considered correlated with toxicity and efficacy. Several designs have been developed to integrate PK information into trial designs. For example, \citet{sun2023pkboin} introduced PKBOIN-12 and TITE-PKBOIN-12, two extensions of BOIN12 that incorporate a continuous PK outcome into Phase I/II trials. Additionally, \citet{takeda2023bayesian} applied PK information within a BOIN-based dose optimization design. Finally, we assumed the same toxicity and efficacy outcomes at both stages. However, this assumption may not hold in some Phase I/II trials. In Phase I, efficacy is typically measured using a short-term outcome, such as an efficacy response during the first few treatment cycles. In contrast, Phase II trials may evaluate long-term efficacy outcomes, such as event-free survival (EFS) or progression-free survival (PFS). The relationship between short-term and long-term efficacy outcomes remains uncertain. As a result, the selected OBD in Phase I based on a short-term efficacy outcome may not necessarily be the same optimal dose based on a long-term efficacy outcome. In future work, we plan to conduct additional simulations to evaluate the performance of existing designs and explore new two-stage designs that incorporate these factors or address more complex scenarios.


\section*{Declaration of Interest Statement}

The authors declare that they have no known competing financial interests or personal relationships that can have appeared to influence the work reported in this paper.

\section*{Data Availability}
This paper does not use any real data. The simulation code will be available on request.

\bibliographystyle{plainnat}
\bibliography{cite}

\clearpage
\begin{longtable}{|l|cccccc|cccccc|}
\caption{True toxicity probabilities, efficacy probabilities, and utility values for each dose level}\label{tab:true_simuvalue}\\
\hline
Category & \multicolumn{12}{c|}{Dose Level}\\
\hline
  & 1 & 2 & 3 & 4 & 5 & 6 & 1 & 2 & 3 & 4 & 5 & 6 \\
\hline
&  \multicolumn{6}{c|}{Scenario 1 (OBD = MTD = 5)} & \multicolumn{6}{c|}{Scenario 2 (OBD = 3, MTD = 4)} \\
Toxicity & 0.05 & 0.10 & 0.15 & 0.22 & \textbf{0.30} & 0.40 & 0.02 & 0.08 & \textbf{0.15} & 0.30 & 0.40 & 0.45 \\
Efficacy & 0.02 & 0.05 & 0.15 & 0.40 & \textbf{0.50} & 0.55 & 0.05 & 0.30 & \textbf{0.45} & 0.50 & 0.55 & 0.58 \\
Utility & 39.2 & 39.0 & 43.0 & 55.2 & \textbf{58.0} & 57.0 & 42.2 & 54.8 & \textbf{61.0} & 58.0 & 57.0 & 56.8 \\  
\hline
& \multicolumn{6}{c|}{Scenario 3 (No OBD, MTD = 3)} & \multicolumn{6}{c|}{Scenario 4 (OBD = MTD = 4)} \\ 
Toxicity & 0.05 & 0.10 & 0.20 & 0.40 & 0.45 & 0.50 & 0.05 & 0.10 & 0.20 & \textbf{0.30} & 0.45 & 0.50 \\
Efficacy & 0.02 & 0.05 & 0.10 & 0.15 & 0.30 & 0.45 & 0.10 & 0.20 & 0.40 & \textbf{0.60} & 0.55 & 0.45 \\
Utility & 39.2 & 39.0 & 38.0 & 33.0 & 40.0 & 47.0 & 44.0 & 48.0 & 56.0 & \textbf{64.0} & 55.0 & 47.0 \\ 
\hline
& \multicolumn{6}{c|}{Scenario 5 (OBD = 3, MTD = 5)} & \multicolumn{6}{c|}{Scenario 6 (No OBD, MTD = 4)} \\
Toxicity & 0.05 & 0.10 & \textbf{0.20} & 0.25 & 0.30 & 0.45 & 0.05 & 0.10 & 0.20 & 0.30 & 0.40 & 0.45 \\
Efficacy & 0.30 & 0.50 & \textbf{0.60} & 0.45 & 0.40 & 0.35 & 0.02 & 0.05 & 0.10 & 0.18 & 0.35 & 0.30\\
Utility & 56.0 & 66.0 & \textbf{68.0} & 57.0 & 52.0 & 43.0 & 39.2 & 39.0 & 38.0 & 38.8 & 45.0 & 40.0 \\ 
\hline
& \multicolumn{6}{c|}{Scenario 7 (OBD = MTD = 4)} & \multicolumn{6}{c|}{Scenario 8 (OBD = 3, MTD = 4)} \\
Toxicity & 0.05 & 0.10 & 0.20 & \textbf{0.30} & 0.40 & 0.45 & 0.05 & 0.15 & 0.20 & \textbf{0.30} & 0.40 & 0.50 \\
Efficacy & 0.05 & 0.20 & 0.40 & \textbf{0.55} & 0.55 & 0.55 & 0.10 & 0.25 & 0.40 & \textbf{0.45} & 0.50 & 0.50 \\
Utility & 41.0 & 48.0 & 56.0 & \textbf{61.0} & 57.0 & 55.0 & 44.0 & 49.0 & 56.0 & \textbf{55.0} & 54.0 & 50.0 \\ 
\hline
& \multicolumn{6}{c|}{Scenario 9 (No OBD, MTD = 4)} & \multicolumn{6}{c|}{} \\
Toxicity & 0.05 & 0.15 & 0.25 & 0.32 & 0.40 & 0.45 & & & & & &\\
Efficacy & 0.02 & 0.08 & 0.15 & 0.20 & 0.20 & 0.20 & & & & & &\\
Utility & 39.2 & 38.8 & 39.0 & 39.2 & 36.0 & 34.0 & & & & & &\\ 
\hline
\end{longtable}

\clearpage
\setlength{\LTleft}{-2cm} 
\footnotesize
\begin{longtable}{|l|c|c|c|c|c|l|c|c|c|c|c|}
\caption{Evaluation of five dose-finding designs for the Phase I part}\label{tab:stage_1_results}\\
\hline
\multicolumn{6}{|c|}{Scenario 1} & \multicolumn{6}{c|}{Scenario 2}\\
\hline
Designs & $p_{rp2d}$ & $p_{rp2d, tox}$ & $p_{et, s1}$ & $n_{tox, s1}$ & $Dur_{s1}$ & Designs & $p_{rp2d}$ & $p_{rp2d, tox}$ & $p_{et, s1}$ & $n_{tox, s1}$ & $Dur_{s1}$ \\
\hline 
BOIN & 69.2\% & 24.0\% & 0.0\% & 7.9 & 24.5 &
BOIN & 63.8\% & 35.6\% & 0.0\% & 12.5 & 24.4 \\
TITE-BOIN & 69.0\% & 26.1\% & 0.0\% & 7.8 & 17.5 & 
TITE-BOIN & 64.3\% & 35.6\% & 0.0\% & 12.2 & 17.4 \\
BF-BOIN & 67.4\% & 22.6\% & 0.0\% & 7.5 & 25.8 & 
BF-BOIN & 66.7\% & 32.8\% & 0.0\% & 13.2 & 25.9 \\
BOIN12 & 65.8\% & 23.4\% & 2.2\% & 4.6 & 52.9 & 
BOIN12 & 77.8\% & 20.3\% & 0.4\% & 6.2 & 52.4\\
TITE-BOIN12 & 66.4\% & 21.6\% & 1.4\% & 5.2 & 33.4 & 
TITE-BOIN12 & 74.9\% & 23.0\% & 0.1\% & 6.5 & 30.4\\
\hline

\multicolumn{6}{|c|}{Scenario 3} & \multicolumn{6}{c|}{Scenario 4}\\
\hline
Designs & $p_{rp2d}$ & $p_{rp2d, tox}$ & $p_{et, s1}$ & $n_{tox, s1}$ & $Dur_{s1}$ & Designs & $p_{rp2d}$ & $p_{rp2d, tox}$ & $p_{et, s1}$ & $n_{tox, s1}$ & $Dur_{s1}$ \\
\hline 
BOIN & NA & 48.8\% & 0.0\% & 20.9 & 24.2 & 
BOIN & 71.9\% & 23.2\% & 0.0\% & 10.0 & 24.3 \\
TITE-BOIN & NA & 49.7\% & 0.0\% & 20.6 & 17.3 & 
TITE-BOIN & 72.5\% & 23.4\% & 0.0\% & 9.7 & 17.3 \\
BF-BOIN & NA & 46.9\% & 0.0\% & 20.8 & 25.6 & 
BF-BOIN & 75.5\% & 18.7\% & 0.0\% & 9.9 & 24.2 \\
BOIN12 & NA & 49.2\% & 19.8\% & 13.5 & 53.1 & 
BOIN12 & 70.5\% & 22.8\% & 0.4\% & 5.2 & 52.2 \\
TITE-BOIN12 & NA & 47.0\% & 16.7\% & 14.3 & 35.8 & 
TITE-BOIN12 & 69.2\% & 24.7\% & 0.6\% & 5.2 & 30.6\\
\hline

\multicolumn{6}{|c|}{Scenario 5} & \multicolumn{6}{c|}{Scenario 6}\\
\hline
Designs & $p_{rp2d}$ & $p_{rp2d, tox}$ & $p_{et, s1}$ & $n_{tox, s1}$ & $Dur_{s1}$ & Designs & $p_{rp2d}$ & $p_{rp2d, tox}$ & $p_{et, s1}$ & $n_{tox, s1}$ & $Dur_{s1}$ \\
\hline 
BOIN & 36.2\% & 15.4\% & 0.0\% & 6.2 & 24.4 & 
BOIN & NA & 34.3\% & 0.0\% & 11.5 & 24.3 \\
TITE-BOIN & 38.0\% & 15.7\% & 0.0\% & 6.0 & 17.5 & 
TITE-BOIN & NA & 33.4\% & 0.0\% & 10.7 & 17.4 \\
BF-BOIN & 42.4\% & 12.6\% & 0.0\% & 5.2 & 26.1 & 
BF-BOIN & NA & 28.5\% & 0.0\% & 11.2 & 25.6 \\
BOIN12 & 85.1\% & 0.7\% & 0.0\% & 0.7 & 51.6 & 
BOIN12 & NA & 32.0\% & 11.6\% & 8.6 & 53.6 \\
TITE-BOIN12 & 83.9\% & 1.3\% & 0.0\% & 0.8 & 27.9 & 
TITE-BOIN12 & NA & 32.8\% & 8.6\% & 8.8 & 36.3 \\
\hline

\multicolumn{6}{|c|}{Scenario 7} & \multicolumn{6}{c|}{Scenario 8}\\
\hline
Designs & $p_{rp2d}$ & $p_{rp2d, tox}$ & $p_{et, s1}$ & $n_{tox, s1}$ & $Dur_{s1}$ & Designs & $p_{rp2d}$ & $p_{rp2d, tox}$ & $p_{et, s1}$ & $n_{tox, s1}$ & $Dur_{s1}$ \\
\hline 
BOIN & 68.6\% & 34.3\% & 0.0\% & 11.5 & 24.3 & 
BOIN & 64.3\% & 32.7\% & 0.0\% & 10.6 & 24.3 \\
TITE-BOIN & 69.6\% & 33.4\% & 0.0\% & 10.7 & 17.4 & 
TITE-BOIN & 64.8\% & 32.2\% & 0.0\% & 10.2 & 17.3 \\
BF-BOIN & 71.9\% & 29.6\% & 0.0\% & 11.6 & 25.9 & 
BF-BOIN & 70.0\% & 27.0\% & 0.0\% & 10.6 & 25.9 \\
BOIN12 & 66.0\% & 29.5\% & 0.6\% & 6.7 & 52.4 & 
BOIN12 & 70.4\% & 22.0\% & 1.0\% & 6.7 & 52.7 \\
TITE-BOIN12 & 66.7\% & 29.4\% & 1.0\% & 6.7 & 30.9 & 
TITE-BOIN12 & 66.8\% & 23.4\% & 0.6\% & 7.0 & 31.5 \\
\hline

\multicolumn{6}{|c|}{Scenario 9} & \multicolumn{6}{|c|}{} \\
\hline
Designs & $p_{rp2d}$ & $p_{rp2d, tox}$ & $p_{et, s1}$ & $n_{tox, s1}$ & $Dur_{s1}$ & \multicolumn{6}{|c|}{} \\
\hline 
BOIN & NA & 26.4\% & 0.0\% & 8.6 & 24.3 & \multicolumn{6}{|c|}{}\\
TITE-BOIN & NA & 27.3\% & 0.0\% & 8.5 & 17.3 & \multicolumn{6}{|c|}{}\\
BF-BOIN & NA & 21.8\% & 0.0\% & 7.9 & 25.6 & \multicolumn{6}{|c|}{}\\
BOIN12 & NA & 21.9\% & 10.7\% & 6.2 & 53.2 & \multicolumn{6}{|c|}{}\\
TITE-BOIN12 & NA & 22.4\% & 8.9\% & 6.4 & 35.9 & \multicolumn{6}{|c|}{}\\
\hline
\end{longtable}

\clearpage
\setlength{\LTleft}{-2.5cm} 
\begin{longtable}{|l|ccc|ccc|ccc|ccc|ccc|}
\caption{Evaluation of fifteen two-stage designs for overall Phase I/II trial performance}\label{tab:overall_results}\\
\hline
Scenario 1 & \multicolumn{3}{c|}{$p_{rp3d}$ (\%)} & \multicolumn{3}{c|}{$p_{et}$ (\%) } & \multicolumn{3}{c|}{$n_{total}$} & \multicolumn{3}{c|}{$n_{tox}$} & \multicolumn{3}{c|}{$Dur$}\\
\hline
Design & TS & BOP2 & TOP & TS & BOP2 & TOP & TS & BOP2 & TOP & TS & BOP2 & TOP & TS & BOP2 & TOP \\
\hline
BOIN & 32.6 & 41.7 & 43.6 & 21.1 & 4.1 & 4.5 & 101.5 & 119.5 & 119.2 & 12.4 & 16.5 & 16.0 & 40.1 & 43.9 & 40.6\\
TITE-BOIN & 30.5 & 44.1 & 42.3 & 22.9 & 5.5 & 5.8 & 100.4 & 119.1 & 118.4 & 12.7 & 16.8 & 16.4 & 32.9 & 36.9 & 33.5\\
BF-BOIN & 28.1 & 40.6 & 43.5 & 24.0 & 6.1 & 5.8 & 111.9 & 129.5 & 129.4 & 12.3 & 15.1 & 14.8 & 41.3 & 45.0 & 41.8\\
BOIN12 & 29.2 & 39.5 & 39.9 & 26.1 & 7.5 & 8.0 & 97.5 & 115.7 & 115.0 & 9.4 & 13.0 & 12.6 & 68.0 & 72.0 & 68.7\\
TITE-BOIN12 & 30.2 & 42.4 & 40.9 & 26.2 & 9.2 & 8.7 & 98.3 & 115.2 & 114.9 & 9.3 & 12.6 & 12.5 & 48.5 & 52.2 & 49.0\\
\hline

Scenario 2 & \multicolumn{3}{c|}{$p_{rp3d}$ (\%) } & \multicolumn{3}{c|}{$p_{et}$ (\%)} & \multicolumn{3}{c|}{$n_{total}$} & \multicolumn{3}{c|}{$n_{tox}$} & \multicolumn{3}{c|}{$Dur$}\\
\hline
Design & TS & BOP2 & TOP & TS & BOP2 & TOP & TS & BOP2 & TOP & TS & BOP2 & TOP & TS & BOP2 & TOP \\
\hline
BOIN & 49.0 & 43.9 & 42.7 & 17.8 & 3.6 & 4.7 & 104.1 & 121.8 & 119.8 & 20.6 & 27.7 & 26.4 & 40.4 & 44.2 & 40.6\\
TITE-BOIN & 48.8 & 45.4 & 46.5 & 16.3 & 3.8 & 5.0 & 104.3 & 121.4 & 120.3 & 19.9 & 26.8 & 26.3 & 33.6 & 37.2 & 33.7\\
BF-BOIN & 49.2 & 48.3 & 46.6 & 14.4 & 3.0 & 2.9 & 121.4 & 137.7 & 136.5 & 20.7 & 26.3 & 25.6 & 42.4 & 45.8 & 42.3\\
BOIN12 & 57.8 & 53.0 & 58.8 & 11.2 & 1.8 & 3.1 & 107.5 & 122.1 & 121.0 & 10.7 & 14.7 & 14.4 & 69.4 & 72.4 & 68.9\\
TITE-BOIN12 & 55.3 & 54.0 & 53.5 & 12.1 & 2.6 & 2.6 & 107.6 & 121.5 & 121.1 & 11.9 & 15.9 & 15.8 & 47.4 & 50.3 & 46.9\\
\hline 

Scenario 3 & \multicolumn{3}{c|}{$p_{rp3d}$ (\%)} & \multicolumn{3}{c|}{$p_{et}$ (\%)} & \multicolumn{3}{c|}{$n_{total}$} & \multicolumn{3}{c|}{$n_{tox}$} & \multicolumn{3}{c|}{$Dur$}\\
\hline
Design & TS & BOP2 & TOP & TS & BOP2 & TOP & TS & BOP2 & TOP & TS & BOP2 & TOP & TS & BOP2 & TOP \\
\hline
BOIN & & & & 97.8 & 84.5 & 80.6 & 81.5 & 97.0 & 99.2 & 29.9 & 38.6 & 38.3 & 34.0 & 38.4 & 36.0\\
TITE-BOIN & & & & 97.6 & 81.1 & 81.9 & 81.3 & 97.0 & 99.3 & 29.6 & 38.7 & 38.3 & 27.0 & 31.4 & 29.0\\
BF-BOIN & & & & 97.5 & 80.5 & 82.0 & 85.6 & 101.1 & 103.4 & 29.0 & 37.3 & 36.8 & 35.2 & 39.7 & 37.3\\
BOIN12 & & & & 97.9 & 85.9 & 85.9 & 66.1 & 78.0 & 79.4 & 22.9 & 31.4 & 31.5 & 61.1 & 65.5 & 63.4\\
TITE-BOIN12 & & & & 97.4 & 83.4 & 83.5 & 67.1 & 80.4 & 80.0 & 23.5 & 33.0 & 32.0 & 43.6 & 48.4 & 45.6\\
\hline

Scenario 4 & \multicolumn{3}{c|}{$p_{rp3d}$ (\%)} & \multicolumn{3}{c|}{$p_{et}$ (\%)} & \multicolumn{3}{c|}{$n_{total}$} & \multicolumn{3}{c|}{$n_{tox}$} & \multicolumn{3}{c|}{$Dur$}\\
\hline
Design & TS & BOP2 & TOP & TS & BOP2 & TOP & TS & BOP2 & TOP & TS & BOP2 & TOP & TS & BOP2 & TOP \\
\hline
BOIN & 37.1 & 59.6 & 58.1 & 18.3 & 4.7 & 5.1 & 102.5 & 120.6 & 119.2 & 14.4 & 18.5 & 17.6 & 40.1 & 43.9 & 40.4\\
TITE-BOIN & 37.7 & 59.6 & 58.0 & 19.2 & 3.8 & 4.0 & 102.2 & 120.4 & 119.9 & 13.7 & 17.9 & 17.6 & 33.1 & 36.9 & 33.6\\
BF-BOIN & 38.4 & 60.1 & 60.6 & 17.4 & 2.7 & 3.5 & 117.7 & 135.2 & 134.2 & 13.0 & 16.3 & 16.0 & 42.0 & 45.6 & 42.2\\
BOIN12 & 37.9 & 60.4 & 56.6 & 20.2 & 3.8 & 4.5 & 101.1 & 118.8 & 118.1 & 9.1 & 12.6 & 12.4 & 67.8 & 71.6 & 68.3\\
TITE-BOIN12 & 35.1 & 58.1 & 54.8 & 23.5 & 5.9 & 5.9 & 99.9 & 117.6 & 117.0 & 9.3 & 13.4 & 13.3 & 45.9 & 49.8 & 46.5\\
\hline

Scenario 5 & \multicolumn{3}{c|}{$p_{rp3d}$ (\%)} & \multicolumn{3}{c|}{$p_{et}$ (\%)} & \multicolumn{3}{c|}{$n_{total}$} & \multicolumn{3}{c|}{$n_{tox}$} & \multicolumn{3}{c|}{$Dur$}\\
\hline
Design & TS & BOP2 & TOP & TS & BOP2 & TOP & TS & BOP2 & TOP & TS & BOP2 & TOP & TS & BOP2 & TOP \\
\hline
BOIN & 25.8 & 29.6 & 28.5 & 12.0 & 1.2 & 1.6 & 106.5 & 122.9 & 122.0 & 8.6 & 11.2 & 10.6 & 41.2 & 44.5 & 41.1\\
TITE-BOIN & 27.8 & 31.0 & 32.4 & 14.3 & 1.0 & 1.7 & 106.0 & 122.8 & 122.1 & 8.2 & 10.8 & 10.7 & 34.0 & 37.5 & 34.1\\
BF-BOIN & 31.5 & 35.5 & 35.6 & 9.9 & 1.1 & 1.0 & 130.1 & 145.1 & 144.1 & 7.4 & 9.2 & 8.7 & 43.3 & 46.3 & 42.8\\
BOIN12 & 56.3 & 64.5 & 64.6 & 2.5 & 0.1 & 0.0 & 116.6 & 124.7 & 124.2 & 0.8 & 0.9 & 0.9 & 70.4 & 71.9 & 68.6\\
TITE-BOIN12 & 58.2 & 61.0 & 64.3 & 3.4 & 0.1 & 0.2 & 115.8 & 124.6 & 124.2 & 1.0 & 1.2 & 1.2 & 46.6 & 48.2 & 44.9\\
\hline

Scenario 6 & \multicolumn{3}{c|}{$p_{rp3d}$ (\%)} & \multicolumn{3}{c|}{$p_{et}$ (\%)} & \multicolumn{3}{c|}{$n_{total}$} & \multicolumn{3}{c|}{$n_{tox}$} & \multicolumn{3}{c|}{$Dur$}\\
\hline
Design & TS & BOP2 & TOP & TS & BOP2 & TOP & TS & BOP2 & TOP & TS & BOP2 & TOP & TS & BOP2 & TOP \\
\hline
BOIN & & & & 85.4 & 49.5 & 51.7 & 83.5 & 104.8 & 106.1 & 19.0 & 26.0 & 25.2 & 34.8 & 40.6 & 37.8\\
TITE-BOIN & & & & 85.6 & 50.5 & 49.9 & 83.6 & 104.4 & 106.0 & 17.7 & 24.2 & 24.1 & 27.9 & 33.5 & 30.8\\
BF-BOIN & & & & 85.0 & 51.1 & 50.7 & 90.0 & 110.0 & 111.5 & 17.4 & 22.4 & 21.9 & 36.3 & 41.7 & 39.0\\
BOIN12 & & & & 88.7 & 58.7 & 59.7 & 73.6 & 90.7 & 91.3 & 16.3 & 23.1 & 22.6 & 62.9 & 68.4 & 65.7\\
TITE-BOIN12 & & & & 87.9 & 58.1 & 58.7 & 74.0 & 91.7 & 91.7 & 16.2 & 23.2 & 22.4 & 45.5 & 50.9 & 48.1\\
\hline

Scenario 7 & \multicolumn{3}{c|}{$p_{rp3d}$ (\%)} & \multicolumn{3}{c|}{$p_{et}$ (\%)} & \multicolumn{3}{c|}{$n_{total}$} & \multicolumn{3}{c|}{$n_{tox}$} & \multicolumn{3}{c|}{$Dur$}\\
\hline
Design & TS & BOP2 & TOP & TS & BOP2 & TOP & TS & BOP2 & TOP & TS & BOP2 & TOP & TS & BOP2 & TOP \\
\hline
BOIN & 30.2 & 47.5 & 48.8 & 23.2 & 3.7 & 4.9 & 100.5 & 121.0 & 119.4 & 19.3 & 26.1 & 25.1 & 39.6 & 44.0 & 40.5\\
TITE-BOIN & 34.3 & 47.9 & 48.3 & 21.6 & 4.2 & 5.1 & 101.0 & 120.2 & 119.4 & 17.9 & 24.3 & 23.9 & 32.8 & 36.9 & 33.5\\
BF-BOIN & 35.3 & 51.8 & 50.8 & 19.3 & 4.1 & 3.6 & 116.2 & 134.2 & 133.2 & 18.6 & 23.6 & 23.4 & 41.7 & 45.5 & 42.1\\
BOIN12 & 33.4 & 49.1 & 47.2 & 22.3 & 4.2 & 4.0 & 100.8 & 119.1 & 118.3 & 13.4 & 18.7 & 18.4 & 67.9 & 71.9 & 68.5\\
TITE-BOIN12 & 33.1 & 47.6 & 48.0 & 21.6 & 4.6 & 5.5 & 99.9 & 118.0 & 117.5 & 13.1 & 18.2 & 17.9 & 46.3 & 50.3 & 46.9\\
\hline

Scenario 8 & \multicolumn{3}{c|}{$p_{rp3d}$ (\%)} & \multicolumn{3}{c|}{$p_{et}$ (\%)} & \multicolumn{3}{c|}{$n_{total}$} & \multicolumn{3}{c|}{$n_{tox}$} & \multicolumn{3}{c|}{$Dur$}\\
\hline
Design & TS & BOP2 & TOP & TS & BOP2 & TOP & TS & BOP2 & TOP & TS & BOP2 & TOP & TS & BOP2 & TOP \\
\hline
BOIN & 39.8 & 43.4 & 43.1 & 20.5 & 3.9 & 4.7 & 101.6 & 121.1 & 119.9 & 17.9 & 23.9 & 23.1 & 39.9 & 44.1 & 40.6\\
TITE-BOIN & 42.6 & 42.6 & 43.2 & 21.1 & 2.9 & 5.2 & 101.7 & 120.7 & 119.8 & 16.8 & 22.7 & 22.1 & 32.9 & 37.0 & 33.6\\
BF-BOIN & 43.9 & 45.0 & 43.9 & 16.2 & 3.3 & 2.7 & 117.7 & 135.4 & 134.8 & 16.2 & 21.0 & 20.7 & 42.0 & 45.7 & 42.3\\
BOIN12 & 47.2 & 47.4 & 45.7 & 20.8 & 4.8 & 5.2 & 102.1 & 119.4 & 118.4 & 11.4 & 15.5 & 14.9 & 68.6 & 72.4 & 68.9\\
TITE-BOIN12 & 43.7 & 44.4 & 46.1 & 21.8 & 3.9 & 5.0 & 101.9 & 119.3 & 118.2 & 11.7 & 16.2 & 15.8 & 47.2 & 51.0 & 47.6\\
\hline

Scenario 9 & \multicolumn{3}{c|}{$p_{rp3d}$ (\%)} & \multicolumn{3}{c|}{$p_{et}$ (\%)} & \multicolumn{3}{c|}{$n_{total}$} & \multicolumn{3}{c|}{$n_{tox}$} & \multicolumn{3}{c|}{$Dur$}\\
\hline
Design & TS & BOP2 & TOP & TS & BOP2 & TOP & TS & BOP2 & TOP & TS & BOP2 & TOP & TS & BOP2 & TOP \\
\hline
BOIN & & & & 86.7 & 43.7 & 47.2 & 83.8 & 106.3 & 107.3 & 13.6 & 18.8 & 18.5 & 34.9 & 41.0 & 38.1\\
TITE-BOIN & & & & 87.0 & 45.6 & 42.5 & 83.1 & 106.5 & 108.1 & 13.4 & 18.6 & 18.6 & 27.6 & 33.9 & 31.2\\
BF-BOIN & & & & 85.2 & 47.3 & 45.3 & 90.0 & 112.3 & 113.8 & 12.1 & 16.1 & 15.8 & 36.3 & 42.3 & 39.5\\
BOIN12 & & & & 90.6 & 54.7 & 53.7 & 73.6 & 92.9 & 94.2 & 10.6 & 15.1 & 15.0 & 62.6 & 68.7 & 66.1\\
TITE-BOIN12 & & & & 89.2 & 54.0 & 54.0 & 74.4 & 93.5 & 94.5 & 10.9 & 15.4 & 15.2 & 45.3 & 51.2 & 48.5\\
\hline
\end{longtable}

\end{document}